\documentclass[conference,letterpaper]{IEEEtran}
\usepackage[cmex10]{amsmath}
\usepackage{amssymb,amsfonts,amsthm}
\usepackage{array,multirow,graphicx}
\usepackage[dvipsnames]{xcolor}
\usepackage{tikz}
\usepackage{mathtools, nccmath}
\usetikzlibrary{positioning}
\usepackage{bbm}
\usepackage{graphicx}
\usepackage{pgfplots} 
\usepackage{mathtools}
\usepackage{derivative}
\pgfplotsset{compat=newest} 
\usepackage{textcomp}
\usepackage{cite,comment,enumitem}
\usepackage{bm}

\usepackage{accents}

\newcommand{\Prob}[1]{\textrm{Pr}\left[#1\right]}
\newcommand{\E}[1]{\textrm{E}\left[#1\right]}
\newcommand{\Eq}[1]{\textrm{E}_q\left[#1\right]}

\theoremstyle{plain}

\theoremstyle{definition}

\usepackage[top = 0.675 in, left = 0.65 in, right = 0.65 in, bottom = 0.85 in]{geometry}

\begin{document}

\title{Polar Coded Quantization for \\ Distributed Source Coding}

\author{%
    \IEEEauthorblockN{M.~Yusuf~\c{S}ener\IEEEauthorrefmark{1}\IEEEauthorrefmark{2}, Gerhard~Kramer\IEEEauthorrefmark{1}, Shlomo Shamai (Shitz)\IEEEauthorrefmark{3}, Ronald B\"ohnke\IEEEauthorrefmark{2}, and Wen Xu\IEEEauthorrefmark{2}}
    \IEEEauthorblockA{\IEEEauthorrefmark{1}%
    School of Computation, Information and Technology, Technical University of Munich, 80333 Munich, Germany}
   \IEEEauthorblockA{\IEEEauthorrefmark{2}%
    Huawei Heisenberg Research Center, Huawei Technologies Duesseldorf GmbH, 80992 Munich, Germany}
   \IEEEauthorblockA{\IEEEauthorrefmark{3}%
    Dept. of Electrical and Computer Engineering, Technion—Israel Institute of Technology, Haifa 3200003, Israel}
}

\maketitle

\begin{abstract}
Scalar quantization and probabilistic shaping are applied to the distributed source coding of Gaussian sources, with mean-square error distortion. A coding scheme with a modulo interval, dithering, and truncated Gaussian shaping is shown to achieve the corner points of the Berger-Tung region. The theory is illustrated by designing short-block-length multilevel 5G polar codes for Wyner-Ziv (WZ) polar coded quantization (PCQ). WZ-PCQ substantially reduces the total distortion compared to separate PCQ of the source blocks.
\end{abstract}

\section{Introduction}
\label{sec:introduction}

Polar codes~\cite{arikan2009channel} are suited to dirty paper (DP) and Wyner-Ziv (WZ) coding \cite{Costa-IT83,Wyner-Ziv-IT76} because they permit practical nested coding for reliability, quantization, and shaping~\cite{Korada-IT10,Sutter-ITW12,Honda-Yamamoto-IT13}. Moreover, their design is based on successive cancellation (SC) decoding, which is compatible with multilevel coded modulation \cite{Seidl-IT13}; see~\cite{Imai:IT77,wachsmann1999multilevel,Boecherer-WC17,Prinz-SPAWC17}. For example, one can apply multilevel coding to the bit levels of amplitude shift keying (ASK) or an equally-spaced scalar quantizer~\cite{Eghbalian-Arani-ISWCS13,Liu-COMM16,Liu-COMM19,Liu-IT21,Liu-ISIT24,Jha-JSAIT22,Mondelli-IT18,Iscan-COMM18,Iscan-IA19,Iscan-TETT20,Wiegart-CL19,Boehnke-COMML20,Runge-ISIT22,Runge-ISIT24,Sener-CL21,Sener-CL24,Sener-ISIT24,Sener-ISIT25}. 

A polar code for multilevel channel coding is referred to as polar coded modulation (PCM) in \cite{Seidl-IT13}. Similarly, we use the term polar coded quantization (PCQ) when using a polar code for multilevel quantization. Special cases include polar lattices \cite{Liu-COMM16,Liu-COMM19,Liu-IT21,Liu-ISIT24,Jha-JSAIT22} and scalar probabilistic shaping with dithering \cite{Sener-ISIT25}. More generally, one may apply multi-dimensional nested lattices~\cite{Zamir-Shamai-ISIT98,zamir2002nested,zamir14,Campello-IT19,Dongbo-IEEEA21} with a modulo operator and dithering.

\subsection{Distributed Source Coding}
\label{subsec:intro-dsc}

Distributed source (DS) coding involves multiple correlated sources, each compressed by an encoder that cannot share information with the others. A decoder reconstructs the sources from the encoder messages. The goal is to design the encoders and decoder to achieve low rates and distortion. The authors of \cite{Berger-78,Tung-thesis-78,Berger-Housewright-IT79} employ WZ coding to establish an achievable rate-distortion (RD) region, also known as the Berger-Tung region. This region is optimal for Gaussian sources and mean-square error (MSE) distortion \cite{wagner2008rate}, which we refer to as the quadratic Gaussian problem.

A closely related setting is the CEO problem, where the encoder sources are noisy versions of a remote source estimated by the decoder \cite{berger1996ceo}. WZ coding is again optimal for the quadratic Gaussian problem \cite{oohama2005rate,prabhakaran2004rate}.

\subsection{Contributions and Organization}
This paper examines two schemes for achieving the corner points of the Berger-Tung region in the two-source Gaussian quadratic problem: quantization without and with a scalar modulo interval. When using polar codes, we label the former scheme PCQ and the latter PCQ-mod \cite{Sener-ISIT25}. Our main contributions are as follows.
\begin{itemize}[leftmargin=*]
    \item We prove that WZ coding with scalar quantization, a modulo interval, dithering, and probabilistic shaping achieves the corner points of the Berger–Tung region.
    \item We design short-block-length multilevel 5G polar codes for PCQ and PCQ-mod and demonstrate their effectiveness.
\end{itemize}
We remark that dithering provides robustness and enables secrecy, but makes code design challenging when the side information is weak; see \cite{Sener-ISIT25}.
Note also that \cite{ghaddar2022channel} applied WZ polar codes to distributed binary sources without multilevel coding. However, multilevel coding is key for practical source coding at moderate to high rates.
A short proof that binary multilevel coding with PCQ and PCQ-mod achieves the desired information rates seems possible by adapting the PCM proof steps in \cite{Runge-ISIT22} to the quantization setting.

This paper is organized as follows. 
Sec.~\ref{sec:preliminaries} reviews notation and theory for WZ coding. Sec.~\ref{sec:dsc} reviews DS coding for the Gaussian quadratic problem. Sec.~\ref{sec:modulo} shows that WZ coding with scalar quantization, a modulo interval, dithering, and probabilistic shaping achieves the corner points of the Berger-Tung region. The theory is illustrated by designing and simulating short-block-length 5G polar codes for PCQ and PCQ-mod. Sec.~\ref{sec:conclusions} concludes the paper.

\section{Preliminaries}
\label{sec:preliminaries}

\subsection{Notation}
\label{subsec:notation}
Bold letters $\bm{x}=(x_1,\dots,x_d)^T$ refer to column vectors.
Let $\bm{1}_d$ be the $d$-dimensional all-ones vector and $I_d$ be the $d\times d$ identity matrix.
The determinant of the square matrix $Q$ is $|Q|$.
Let $\bm{x}\circ\bm{y}$ be the Hadamard (entry-by-entry) product of $\bm{x}$ and $\bm{y}$. Define the vector modulo operator
\begin{align}
    \bm{x}\ \textrm{mod}\ \bm{A} = \bm{x} - \bm{k} \circ\bm{A} 
\end{align}
where $\bm{A}$ has positive entries and $k_i$ is the unique integer for which $x_i-k_i A_i \in [-A_i/2,A_i/2)$ for $i=1,\dots,d$.

Upper- and lowercase letters typically denote random variables (RVs) and their realizations, e.g., $X$ and $x$. $P_X$ and $p_X$ are a probability mass function and density, respectively. We remove subscripts if the argument is the lowercase of the RV, e.g., $p(x)=p_X(x)$.
$\E{\bm{X}}$ and the random vector $\E{\bm{X}|\bm{Y}}$ are the expectations of $\bm{X}$ without and with conditioning on $\bm{Y}$, respectively. The corresponding covariance matrices are
\begin{align}
    Q_{\bm{X}} & = \E{(\bm{X}-\E{\bm{X}})(\bm{X}-\E{\bm{X}})^T} 
    \label{eq:cov-matrix} \\
    Q_{\bm{X}|\bm{Y}}
    & = \E{(\bm{X}-\E{\bm{X}|\bm{Y}})(\bm{X}-\E{\bm{X}|\bm{Y}})^T} .
    \label{eq:cond-cov-matrix}
\end{align}
For scalars, we write 
\eqref{eq:cov-matrix} and \eqref{eq:cond-cov-matrix} as $\sigma_x^2$ and  $\sigma_{x|y}^2$.
The notation $h(\bm{X})$, $I(\bm{X};\bm{Y})$, $I(\bm{X};\bm{Y}|\bm{Z})$ refers to the differential entropy of $\bm{X}$ and the mutual information of $\bm{X}$ and $\bm{Y}$ without and with conditioning on $\bm{Z}$, respectively. We write $\Eq{f(X)}:=\int_{\mathbb R} q(x) f(x)\, dx$ and $h_q(X):=\Eq{-\log q(X)}$.

\subsection{Wyner-Ziv Rates}
\label{subsec:WZ-rates}
Let $(\bm{X}_{i},\bm{Y}_{i})$, $i \in \{1\, \dots\, n\}$, be independent and identically distributed (i.i.d.) samples from $p_{\bm{X},\bm{Y}}$. The encoder compresses the source block $\bm{X_1},\, \dots\, \bm{X}_n$ while only the decoder has access to $\bm{Y_1},\, \dots\, \bm{Y}_n$. The RD function of this problem is
\begin{align}
    R_{\text{WZ}}(\mathcal D) = \min_{\E{d(\bm{X},\hat{\bm{X}})}\le \mathcal D} \left( I(\bm{U};\bm{X}) - I(\bm{U};\bm{Y}) \right )
    \label{eq:WZ-RD-function}
\end{align}
where the minimization is over all $\bm{U}$ such that $\bm{U}\leftrightarrow \bm{X}\leftrightarrow \bm{Y}$ forms a Markov chain, and all functions $f(\cdot)$ such that $\hat{\bm{X}} = f(\bm{U},\bm{Y})$. The Markov chain relation gives
\begin{align}
    I(\bm{U};\bm{X}) - I(\bm{U};\bm{Y})
    = I(\bm{U} ; \bm{X} | \bm{Y}).
    \label{eq:R-diff-cond}
\end{align}

Remarkably, $R_{\text{WZ}}(\mathcal D)$ is the same as if the encoder also has $\bm{Y}$ if $\bm{X},\bm{Y}$ are jointly Gaussian and $d(\bm{x},\hat{\bm{x}})=\|\bm{x}-\hat{\bm{x}}\|^2$ (the quadratic Gaussian problem). For example, consider the scalar problem and choose $U=X+\check{Z}$ where $\check{Z}$ is independent of $(X,Y)$. The best estimate of $X$ is
\begin{align}
    \hat{X} = \E{X|U,Y}
    = \frac{\sigma_{x|y}^2\, U + \sigma_{\check{z}}^2\, \E{X|Y}}{\sigma_{x|y}^2+\sigma_{\check{z}}^2}.
    \label{eq:WZ-estimate}
\end{align}
If $0<\mathcal{D}< \sigma_{x|y}^2$, choose the description noise variance as
\begin{align}
    \sigma_{\check{z}}^2
    = \frac{\sigma_{x|y}^2 \mathcal D}{\sigma_{x|y}^2 - \mathcal D} 
    \implies
    \mathcal{D} = \frac{\sigma_{x|y}^2 \sigma_{\check{z}}^2}{\sigma_{x|y}^2 + \sigma_{\check{z}}^2}
    \label{eq:WZ-variances}
\end{align}
This choice gives $\E{(X-\hat{X})^2}=\mathcal{D}$ and
\begin{align}
    I(U;X) - I(U;Y)
    & = \frac{1}{2}\log(\sigma_{x|y}^2\, \big/ \mathcal D).
    \label{eq:WZ-rate-scalar}
\end{align}

\subsection{WZ Coding with Scalar Modulo Operator and Dithering}
\label{subsec:WZ-coding}
\begin{figure}[!t]
\centering
\begin{tikzpicture}[scale=0.75, every node/.style={scale=0.75}]
\coordinate (a) at (2.5,3.2);
\node[above = 1pt of a] {$\bm{x}$};

\coordinate (b) at (3.3,0.2);
\node[above = 0.5pt of b] {$\bm{d}$};

\coordinate (c) at (5.1,2);
\node[above = 1pt of c] {$w$};

\draw [thick](1,0.2) -- (7.8,0.2);
\draw [-latex,thick](3,0.2) -- (2,0.2);
\draw [-latex,thick](3,0.2) -- (4.2,0.2);
\draw [-latex,thick](1,0.2) -- (1,1.86);
\draw [-latex,thick](6.2,0.2) -- (6.2,1.68);
\draw [-latex,thick](3,0.7) -- (3,0.2);

\node[thick] (ee) at (2.99,3.2){$\bigoplus$};

\draw [thick](1,3.2)--(1,2.81);
\draw [-latex,thick](1,2.5)--(1,2.15);
\draw [-latex,thick](2.82,3.2)--(2,3.2);
\draw [thick](2.1,3.2)--(1,3.2);

\draw [-latex,thick](6.2,3.2)--(3.16,3.2);
\draw [-latex,thick](3.16,3.2)--(5,3.2);

\draw [-latex,thick](6.2,3.2) -- (6.2,2.3);

\draw [-latex,thick](3,3.9) -- (3,3.38);
\draw [-latex,thick](4.1,3.7) -- (4.1,3.2);

\draw [-latex,thick](1.2,2) -- (1.56,2);
\draw [-latex,thick](2.8,2) -- (3.4,2);
\node[thick] (aaa) at (1,2){$\bigoplus$};

\draw [-latex,thick](4.83,2) -- (5.48,2);
\coordinate (c2) at (4.6,1.76);

\node[thick] (aa) at (7.8,2){$\bigoplus$};

\draw [thick](6.2,3.2) -- (8,3.2);
\draw [-latex,thick](7.8,3.2) -- (7.8,2.81);
\draw [-latex,thick](7.8,2.5) -- (7.8,2.15);
\draw [-latex,thick](7.8,0.2) -- (7.8,1.86);
\draw [-latex,thick](6.91,2) -- (7.62,2);


\node[thick] (bb) at (10.4,2){$\bigotimes$};
\node[thick] (cc) at (7.8,2.65){$\bigotimes$};
\node[thick] (ccc) at (1,2.65){$\bigotimes$};

\draw [-latex,thick](7.99,2) -- (8.35,2);

\draw [-latex,thick](9.6,2) -- (10.22,2);
\draw [-latex,thick](10.58,2.01) -- (10.96,2.01);

\node[thick] (dd) at (11.1,2){$\bigoplus$};

\draw [-latex,thick](8,3.2) -- (9.1,3.2);
\draw [thick](9,3.2) -- (11.1,3.2);
\draw [-latex,thick](11.1,3.2) -- (11.1,2.16);

\draw [-latex,thick](11.28,2) -- (11.68,2);

\coordinate (d) at (4.1,2.6);
\node[above = 1pt of d] {$\bm{y}$};
\coordinate (e) at (7.15,2);
\node[above = 1pt of e] {$\bm{u}$};
\coordinate (f) at (8.13,2.4);
\node[above = 2pt of f] {$\alpha$};

\coordinate (g) at (8,2);
\node[above = 1pt of g] {-};

\coordinate (h) at (8,1.5);
\node[above = 1pt of h] {-};

\coordinate (i) at (10.4,2.1);
\node[above = 1pt of i] {$\alpha$};
\node[below = 4pt of i, xshift=-0.5cm] {$\bm{z}'$};

\coordinate (ii) at (1,2.7);
\node[right = 2pt of ii] {$\alpha$};
\coordinate (ij) at (3.1,2);
\node[above = 1pt of ij] {$\bm{x}'$};

\coordinate (j) at (11.8,1.76);
\node[above = 1pt of j] {$\hat{\bm{x}}$};

\coordinate (k) at (3,3.83);
\node[above = 1pt of k] {$\bm{z}$};

\node [draw, shape=rectangle,thick, minimum width= 1.0cm, minimum height=0.6cm, text width=0.6cm, anchor=center] at (4.1,4) {$p(y)$};
\node [draw, shape=rectangle,thick, minimum width= 1.0cm, minimum height=0.6cm, text width=0.6cm, anchor=center] at (3,1.02) {$p(d)$};
\node [draw, shape=rectangle,thick, minimum width= 1cm, minimum height=0.6cm, text width=1cm, anchor=center] at (2.16,2) {mod~$A$};
\node [draw, shape=rectangle,thick, minimum width= 1cm, minimum height=0.6cm, text width=1cm, anchor=center] at (8.98,2) {mod~$A$};
\node [draw, shape=rectangle,thick, minimum width= 1.2cm, minimum height=0.6cm, text width=1.2cm, anchor=center] at (4.1,2) {Encoder};
\node [draw, shape=rectangle,thick, minimum width= 1.2cm, minimum height=0.6cm, text width=1.2cm, anchor=center] at (6.2,2) {Decoder};

\end{tikzpicture}
\caption{WZ coding with a modulo operator and dithering.}
\label{fig:DPC-dither}
\end{figure}
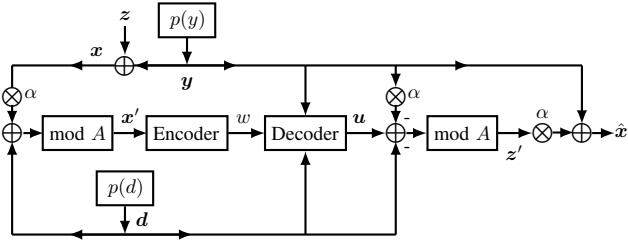

Consider a source signal $X=Y+Z$ where $Z$ is Gaussian and independent of $Y$, i.e., we have $\sigma_{x|y}^2=\sigma_z^2$. The side information $Y$ is only known to the decoder. Fig.~\ref{fig:DPC-dither} shows the WZ coding scheme analyzed in  \cite{Sener-ISIT25}, where the encoder and decoder share a dither $D$ statistically independent of $(X,Z)$ and uniformly distributed in $[-A/2, A/2)$ for $A > 0$. The inflation factor is $\alpha=\sqrt{1-\sigma_d^2/\sigma_z^2}$, where $\sigma_d^2$ is a target distortion; cf. \cite[p.~1260]{zamir2002nested}.

Dithering the source and side information signals, we have
\begin{align}
    (X',Y')^T = \left( \alpha\, (X,Y)^T + D\cdot\bm{1}_2 \right)\ \textrm{mod}\ A\cdot\bm{1}_2 .
    \label{eq:derived-source}
\end{align}
The derived source $X'$ and side information $Y'$ are uniform over $[-A/2, A/2)$ and independent of $X$ and $Y$, respectively. The source $X'$ is quantized via $U$ with discrete alphabet $\mathcal{U}$. The decoder can recover $U$ if the rate satisfies
\begin{align}
    R = I(U;X') - I(U;Y')
    \label{eq:WZ-rate}
\end{align}
where $U\leftrightarrow X'\leftrightarrow Y'$ forms a Markov chain. 

Let $\sigma_d^2$ be a distortion parameter satisfying $0 < \sigma_d^2 < \sigma_z^2$.
Define the encoder and decoder noise
\begin{align}
    \tilde{Z} & = (U - X')\ \textrm{mod}\ A
    \label{eq:Ztilde} \\
    Z' & = (U - Y')\ \textrm{mod}\ A
    = \big( \tilde Z  + \alpha Z \big)\ \textrm{mod}\ A
   \label{eq:Zprime}
\end{align}
where $U$ is selected based on $X'$. Thus, $U\leftrightarrow X'\leftrightarrow Y'$ forms a Markov chain, as required. The decoder outputs the reconstruction
\begin{align}
    \hat{X} = Y + \alpha Z'.
    \label{eq:reconstruction}
\end{align}

Now consider the $M$-ASK quantization alphabet
\begin{equation}
 \mathcal{U} = \big\{ -A/2 + (k+1/2)\kappa \ \big\}_{k=0}^{M-1}
 \label{eq:constellation}
\end{equation}
where $A\ge0$ and $\kappa = A/M$ is the ASK spacing.
Let $q(\tilde z)$ be a shaping density on $\tilde{z} \in [-A/2,A/2)$ such that
\begin{equation}
    P(u|x') = \frac{q\big((u-x')\ \text{mod}\ A\big)}{\sum_{v\in\mathcal U} q\big((v-x')\ \text{mod}\ A\big)}, \quad u \in \mathcal U .
\label{eq:q_shaping}
\end{equation}
The continuously uniform $X'$ induces a discrete uniform $U$, and $U,\tilde{Z},Z$ are mutually statistically independent \cite{Sener-ISIT25}. Consider truncated Gaussian shaping with \cite[eq.~(12)]{Sener-CL24}
\begin{equation}
q(\tilde{z}) = \frac{e^{-\tilde{z}^2 / (2\sigma_d^2)}}{c \cdot \sqrt{2\pi\sigma_d^2}}, \quad \tilde{z} \in [-A/2,A/2)
\label{eq:TG-shaping}
\end{equation}
where $c=1-2Q(A/(2\sigma_d))$. For $M\to\infty$ and $A\to\infty$, one achieves the WZ-RD region \cite{Sener-ISIT25}:
\begin{align}
    \mathcal{D} &= \E{\big(X - \hat X\big)^2} \to \sigma_d^2 \\
    R &= I(U;X') - I(U;Y') \to \frac{1}{2}\log(\sigma_{z}^2 / \sigma_{d}^2). 
\end{align}

\section{Review of DS Coding}
\label{sec:dsc}

Consider a source that outputs $(\bm{X}_1,\bm{X}_2)$ where the pairs $(X_{1,i},X_{2,i})$, $i=1,\dots,n$, are i.i.d. with distribution $P(x_1,x_2)$ or density $p(x_1,x_2)$. There are two encoders and one decoder. Encoder $l$, $l=1,2$, compresses the string $X_l^n$ to the $n R_l$ bits $W_l$, and the decoder reconstructs the pair $(\hat{\bm X}_1,\hat{\bm X}_2)$ from $(W_1,W_2)$.
Let $d_1(.)$ and $d_2(.)$ be two non-negative, real-valued distortion functions. The goal of DS coding is to design the encoders and decoder so that the empirical distortions
\begin{align}
    \Delta_l & := \frac{1}{n} \sum\nolimits_{i=1}^n d_l(X_{l,i},\hat X_{l,i})
\end{align}
satisfy $\E{\Delta_l}\le \mathcal D_l$ for $l=1,2$.

An achievable region is obtained with separate WZ encoding and joint decoding \cite{Berger-78,Tung-thesis-78,Berger-Housewright-IT79}. The resulting Berger-Tung region is the union of $(R_1,R_2,\mathcal D_1,\mathcal D_2)$ satisfying
\begin{subequations}
\begin{align}
    R_1 & \ge I(X_1 ; U_1 | U_2, T) \\
    R_2 & \ge I(X_2 ; U_2 | U_1, T) \\
    R_1+R_2 & \ge I(X_1, X_2 ; U_1, U_2 | T)
\end{align}
\end{subequations}
and 
\begin{align}
    \mathcal D_1 \ge \E{d_1(X_1,\hat{X}_1)}, \quad
    \mathcal D_2 \ge \E{d_2(X_2,\hat{X}_2)} 
\end{align}
where the joint distribution of RVs factors as
\begin{align}
    p(t)\, p(x_1,x_2)\, p(u_1|x_1,t)\, p(u_2|x_2,t)
    \label{eq:TS-long-Markov-chain}
\end{align}
and $\hat X_l=f_l(U_1,U_2,T)$ for some functions $f_l(.)$, $l=1,2$.

\subsection{Scalar Quadratic Gaussian Problem}
\label{subsec:dsc-Gauss}

Consider jointly Gaussian $X_1,X_2$ with zero mean, covariance matrix $Q_{\bm{X}}$, and 
$d_l(\bm{x},\hat{\bm{x}})=\|\bm{x}-\hat{\bm{x}}\|^2$
for $l=1,2$. To encode, set $T=0$ and 
\begin{align}
    \bm{U} = \bm{X} + \check{\bm{Z}}
    \label{eq:dsc-Gauss-U}
\end{align}
where $\check{\bm{Z}}$ is independent of $\bm{X}$ and has independent Gaussian entries. The reconstructions are $\hat{\bm {X}}=\E{\bm{X}| \bm{U}}$. 

The Berger-Tung region is the union over all $(\sigma_{\check z_1}^2,\sigma_{\check z_2}^2)$ of the $(R_1,R_2,\mathcal D_1,\mathcal D_2)$ satisfying \cite[Eq. (12.3)]{ElGamal-Kim-11}
\begin{subequations}
\begin{align}
    & R_1 \ge \frac{1}{2} \log \frac{|Q_{\bm{U}}|}{\sigma_{u_2}^2 \sigma_{\check z_1}^2 }, \quad
    R_2 \ge \frac{1}{2} \log \frac{|Q_{\bm{U}}|}{\sigma_{u_1}^2 \sigma_{\check z_2}^2} 
    \label{eq:BT-Gauss-rates12}
    \\
    & R_1+R_2 \ge \frac{1}{2} \log \frac{|Q_{\bm{U}}|}{\sigma_{\check z_1}^2 \sigma_{\check z_2}^2}
    \label{eq:BT-Gauss-sum-rate}
\end{align}
\label{eq:BT-Gauss-rates}
\end{subequations}
where 
\begin{align}
    \mathcal D_1
    \geq \frac{|Q_{X_1,U_2}|\, \sigma_{\check z_1}^2}{|Q_{\bm{U}}|}, \quad
    \mathcal D_2
    \geq \frac{|Q_{X_2,U_1}|\, \sigma_{\check z_2}^2}{|Q_{\bm{U}}|} .
    \label{eq:BT-Gauss-D}
\end{align}
The distortion terms can be rewritten as \cite[p. 301]{ElGamal-Kim-11}
\begin{subequations}
\begin{align}
    \mathcal D_1 & 
    \geq  \frac{\sigma_ {x_1}^2 \big( \sigma_{x_2}^2(1-\rho^2)+\sigma_{\check z_2}^2 \big)\, \sigma_{\check z_1}^2}{|Q_{\bm{U}}|} 
    \label{eq:BT-Gauss-D1}
    \\
    \mathcal D_2 & 
    \geq \frac{\sigma_ {x_2}^2 \big( \sigma_{x_1}^2(1-\rho^2)+\sigma_{\check z_1}^2 \big)\, \sigma_{\check z_2}^2}{|Q_{\bm{U}}|} .
    \label{eq:BT-Gauss-D2}
\end{align}
\label{eq:BT-Gauss-D12}
\end{subequations}
where $\rho = \E{X_1 X_2}/(\sigma_{x_1}\sigma_{x_2})$ is the correlation coefficient. 

\begin{figure}[!t]
\centering
\input{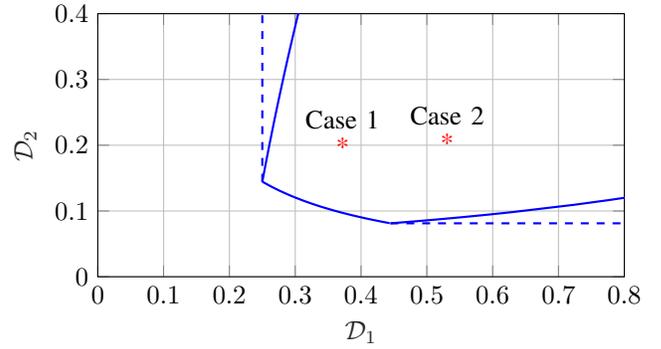}
\caption{Distortion region for a Gaussian source with covariance matrix \eqref{eq:QX-example} and $(R_1,R_2)=(1,2)$. The markers show average distortion pairs achieved with 5G polar codes for $n=256$ source symbols; see Sec.~\ref{subsec:DS-PCQ}.
}
\label{fig:BT-region}
\end{figure}

Fig.~\ref{fig:BT-region} shows the achievable distortions \eqref{eq:BT-Gauss-D12} for $(R_1,R_2)=(1,2)$ by varying the pairs $(\sigma_{\check z_1}^2,\sigma_{\check z_2}^2)$ for the source
\begin{align}
    Q_{\bm{X}} = \begin{pmatrix}
        2.5 & 2 \\ 2 & 2.5
    \end{pmatrix}
    \label{eq:QX-example}
\end{align}
and checking if the rate bounds \eqref{eq:BT-Gauss-rates} are satisfied. The solid curve plots the pairs $(\mathcal D_1,\mathcal D_2)$ for which equality holds in \eqref{eq:BT-Gauss-D12}; the dashed lines complete the region boundary. The markers show distortion pairs achieved with $n=256$ in Sec.~\ref{subsec:DS-PCQ}.

\subsection{Successive Decoding}
\label{subsec:dsc-successive}

The Berger-Tung tuples $(R_1,R_2,\mathcal D_1,\mathcal D_2)$ can be achieved with joint decoding, but separate decoding is easier to implement. A standard approach targets the region's corner points, and uses either time-sharing, offset encoding, or rate splitting to achieve the other points; see \cite{Chen-Berger-IT08}, \cite[Sec.~2.2]{Kim-Kramer-22}.

More precisely, the factorization \eqref{eq:TS-long-Markov-chain} implies that, for each $T=t$, we have the long Markov chain
$U_1 \leftrightarrow X_1 \leftrightarrow X_2 \leftrightarrow U_2$ and therefore $U_1 \leftrightarrow X_1 \leftrightarrow X_2 \leftrightarrow U_2$ and
\begin{align}
    I(X_1 ; U_1 | U_2, T) & = I(X_1 ; U_1 | T) - I(U_1 ; U_2 | T) \label{eq:dsc-WZ-mi1} \\
    I(\bm{X} ; \bm{U} | T) & = I(X_2 ; U_2 | T) + I(X_1 ; U_1 | U_2, T). \label{eq:dsc-WZ-mi12}
\end{align}
These expressions motivate using successive decoding to operate near the corner point 
\begin{align}
     (R_1,R_2) = \big( I(X_1 ; U_1 | U_2, T),\, I(X_2 ; U_2 | T) \big).
\end{align}
Here, encoder~2 uses Shannon-RD coding at rate $I(X_2 ; U_2 | T)$ and encoder~1 uses WZ coding at rate \eqref{eq:dsc-WZ-mi1}, treating $U_2$ as side information; see \cite{Chen-Berger-IT08}. The decoder obtains $\bm{U}$ and computes the same $\hat{\bm{X}}$ as in joint decoding. Swapping the indexes 1 and 2, one can also operate near the corner point
\begin{align}
     (R_1,R_2) = \big( I(X_1 ; U_1 | T),\, I(X_2 ; U_2 | U_1, T) \big).
\end{align}
All other rate pairs may be obtained by time-sharing.

So consider the Gaussian source with
\begin{align}
    R_1 & = I(U_1 ; X_1 | U_2) 
    = \frac{1}{2} \log \frac{|Q_{\bm{U}}|}{\sigma_{u_2}^2 \sigma_{\check z_1}^2}
    \label{eq:dsc-corner-1} \\
    R_2 & = I(X_2 ; U_2) = \frac{1}{2} \log \frac{\sigma_{u_2}^2}{\sigma_{\check z_2}^2} .
    \label{eq:dsc-corner-2}
\end{align}
Encoder 2 can compress $X_2$ with distortion $\sigma_{d_2}^2$ near rate $R_2=\frac{1}{2}\log(\sigma_{x_2}^2/\sigma_{d_2}^2)$, so using \eqref{eq:dsc-corner-2} we have (see \eqref{eq:WZ-variances})
\begin{align}
    \sigma_{d_2}^2 = \frac{\sigma_{x_2}^2 \sigma_{\check z_2}^2}{\sigma_{x_2}^2 + \sigma_{\check z_2}^2}
    \implies
    \sigma_{\check z_2}^2 = \frac{\sigma_{x_2}^2 \sigma_{d_2}^2}{\sigma_{x_2}^2 - \sigma_{d_2}^2} .
    \label{eq:dsc-sigma-d2}
\end{align}

Next, consider $U_2$ as side information at the decoder. Encoder 1 can compress $X_1$ with distortion $\mathcal D_1$ in \eqref{eq:BT-Gauss-D1} near the WZ rate (see \eqref{eq:BT-Gauss-D})
\begin{align}
    R_1 = \frac{1}{2} \log \frac{\sigma_{x_1|u_2}^2}{\mathcal D_1}
    = \frac{1}{2} \log \frac{|Q_{\bm{U}}|}{ \sigma_{u_2}^2 \sigma_{\check z_1}^2}
    \label{eq:dsc-sigma-d1}
\end{align}
which is the same as \eqref{eq:dsc-corner-1}. Define $U_1=X_1+\check Z_1$ where $\check Z_1$ is independent of $(X_1,X_2,\check Z_2)$ so $\bm{U}$, $\bm{X}$, $\check{\bm{Z}}$ are related as in \eqref{eq:dsc-Gauss-U}. Moreover, one can identify $(X_1,U_2,U_1,\check Z_1)$ as the $(X,Y,U,\check Z)$ in \eqref{eq:WZ-estimate}. The decoder can thus compute $\bm{U}$ and $\hat{\bm X} = \E{\bm{X}|\bm{U}}$ with distortions $\mathcal D_1,\mathcal D_2$ in \eqref{eq:BT-Gauss-D}.

\section{DS Coding with a Modulo Operator}
\label{sec:modulo}

We show that scalar quantization with a modulo interval, dithering, and probabilistic shaping achieves the corner points of the Berger-Tung region. Consider successive decoding for dithered modulo intervals. Encoder $l$, $l=1,2$, uses the interval $[-A_l/2,A_l/2)$ and $M_l$-ASK with quantizer alphabet
\begin{equation}
 \mathcal{U}_l = \big\{-A_l/2 + (k+1/2)\kappa_l\ \big\}_{k=0}^{M_l-1}
 \label{eq:dsc-constellation}
\end{equation}
where $\kappa_l = A_l/M_l$ is the ASK spacing. Encoder $l$ computes
\begin{align}
    X_l' = (\alpha_l X_l + D_l)\ \textrm{mod}\ A_l
    \label{eq:dsc-Xsp}
\end{align}
with inflation factors $\alpha_l$ and independent dithers $D_1,D_2$. Encoder $l$ selects $U_l$ using $X_l'$ and a truncated Gaussian shaping density $q(.)$ with parameter $\sigma_{d_l}^2$, which defines $c_l$. It then sends the message $W_l$. Define the encoder noise as
\begin{align}
    \tilde Z_l = (U_l-X_l')\ \textrm{mod}\ A_l.
\end{align}

The decoder obtains $\bm{U}$ from $W_1,W_2$, and uses $\bm{U}$ and $\bm{D}$ to compute a pair $Y_1,Y_2$ as defined below.
It then sets $Y_l'=(\alpha_l Y_l + D_l)\ \textrm{mod}\ A_l$ for $s=1,2$ and
\begin{align}
    Z_l' & = \big( U_l - Y_l' \big)\ \textrm{mod}\ A_l \nonumber \\ &
    = (\tilde Z_l + \alpha_l Z_l)\ \textrm{mod}\ A_l
    \label{eq:Zs-prime}
\end{align}
where $Z_l = X_l - Y_l$.

\subsection{Encoder 2}
Consider the corner point \eqref{eq:dsc-corner-2}. Encoder 2 compresses $X_2$ with distortion $\sigma_{d_2}^2$ in \eqref{eq:dsc-sigma-d2} and sets
\begin{align}
    \alpha_2 = \sqrt{1-\sigma_{d_2}^2 \big/ \sigma_{x_2}^2} = \sigma_{x_2}/\sigma_{u_2}.
    \label{eq:alpha2}
\end{align}
Moreover, set $Y_2:=0$ so $Y_2'=D_2$ and
\begin{align}
    Z_2' = (\tilde Z_2 + \alpha_2 X_2)\ \textrm{mod}\ A_2.
\end{align}
Encoder 2 uses the shaping density \eqref{eq:q_shaping} with $\sigma_d^2=\sigma_{d_2}^2$, so $\tilde Z_2$ has variance (see (see~\cite[eq.~(16)]{Sener-CL24}))
\begin{align}
    P_{q,\tilde Z_2}
    = \sigma_{d_2}^2 \left[ 1 - \frac{A_2 e^{-A_2^2/(8\sigma_{d_2}^2)}}{c_2 \sqrt{2 \pi \sigma_{d_2}^2}} \right] .
    \label{eq:Ztilde-variance-2}
\end{align}
The value \eqref{eq:Ztilde-variance-2}  is approximately $\sigma_{d_2}^2$ for large $A_2$ and $M_2$. Similarly, $Z_2'$ has (see~\cite[eq.~(42)]{Sener-ISIT25})
\begin{align}
    \E{(Z_2')^2} \le \sigma_{x_2}^2 + \left(P_{q,\tilde Z_2} \big/ d_\textrm{min,2} - \sigma_{d_2}^2 \right)
    \label{eq:Zprime-variance-bound2-2}
\end{align}
where $d_\textrm{min,2}$ is defined as in~\cite[eq.~(40)]{Sener-ISIT25}, and the right hand side (RHS) of \eqref{eq:Zprime-variance-bound2-2}
is approximately $\sigma_{x_2}^2$ for large $A_2$ and $M_2$. 

\subsection{Encoder 1}
Encoder 1 mimics the approach in Sec.~\ref{subsec:WZ-coding} and chooses the side information $Y_1$ so $X_1=Y_1+Z_1$ where $Y_1$ and $Z_1$ are uncorrelated. An appropriate choice is 
\begin{align}
    Y_1 = \frac{\E{X_1 Z_2'}}{\E{(Z_2')^2}} \cdot Z_2', \quad Z_1 = X_1 - Y_1.
\end{align}
We approximate and set
\begin{align}
    Y_1 & := \frac{\E{X_1 \big( \alpha_2 X_2 + \tilde Z_2 \big)}}{\sigma_{x_2}^2} \cdot Z_2'
    = \frac{\rho\,\sigma_{x_1}}{\sigma_{u_2}} \cdot Z_2' .
\end{align}
Moreover, set
\begin{align}
    \alpha_1 &= \sqrt{1-\mathcal D_1 \big/ \sigma_{x_1|u_2}^2}
    \overset{(a)}{=} \sqrt{\frac{|Q_{X_1,U_2}|}{|Q_{\bm{U}}|}} 
    \label{eq:alpha1}
\end{align}
where step $(a)$ follows because
$\sigma_{u_2|x_1}^2 = \sigma_{u_2}^2 - \rho^2 \sigma_{x_2}^2$.
Let encoder 1 use the shaping density \eqref{eq:q_shaping} with $\sigma_{d_1}^2=\mathcal D_1$. The decoder can now compute (see \eqref{eq:Zs-prime})
\begin{align}
    Z_1' & = \left(\tilde Z_1 + \alpha_1 Z_1 \right)\ \textrm{mod}\ A_1 \\
    \hat X_1 & = Y_1 + \alpha_1 Z_1'
    \label{eq:dsc-sc-hatX1}
\end{align}
where $\tilde Z_1=(U_1-X_1')\ \textrm{mod}\ A_1$ and $Z_1=X_1-Y_1$ and
\begin{align}
    X_1 -\hat X_1 = Z_1 - \alpha_1 Z_1'.
\end{align}
We bound 
\begin{align}
    & \sigma_{z_1}^2 = \E{\left( X_1 - \gamma_1 Z_2' \right)^2} \nonumber \\
    & \le \sigma_{x_1|u_2}^2
    + 2 \gamma_1 A_2 \E{X_1 I_2} 
    + \gamma_1^2 \left( \frac{P_{q,\tilde Z_2}}{d_\textrm{min,2}} - \sigma_{d_2}^2 \right)
    \label{eq:Zprime-variance-bound2-1}
\end{align}
which follows by \eqref{eq:Zprime-variance-bound2-2} and
\eqref{eq:Zs-prime} with
$Z_2'=\tilde Z_2+\alpha_2 Z_2 - I_2 A_2$
and $\sigma_{x_1|u_2}^2=\sigma_{x_1}^2(1-\alpha_2^2 \rho^2)$. We also have (see~\cite[eq.~(43)]{Sener-ISIT25})
\begin{align}
    \E{Z_1 Z_1'}
    & = \alpha_1 \sigma_{z_1}^2 - A_1 \E{I_1 Z_1}.
    \label{eq:cross-correlation-1}
\end{align}
Finally, consider (see~\cite[eq.~(42)]{Sener-ISIT25})
\begin{align}
    \E{(Z_1')^2} \le \sigma_{z_1}^2 + \left( \frac{P_{q,\tilde Z_1}}{d_\textrm{min,1}} - \frac{\sigma_{z_1}^2}{\sigma_{x_1|u_2}^2} \mathcal D_1 \right)
    \label{eq:Zprime-variance-bound2-3}
\end{align}
where $P_{q,\tilde Z_1}$ and $d_\textrm{min,1}$ are defined as for encoder 2. Combining the results, we have (see~\cite[eq.~(44)]{Sener-ISIT25})
\begin{align}
    & \E{ \big( X_1 - \hat X_1 \big)^2}
    = \sigma_{z_1}^2
    - 2 \alpha_1\, \E{Z_1 Z_1'}
    + \alpha_1^2\,\E{(Z_1')^2} \nonumber \\
    & \overset{(a)}{\le}
    \frac{\sigma_{z_1}^2}{\sigma_{x_1|u_2}^2} \mathcal D_1
    + 2 \alpha_1 A_1 \, \E{I_1 Z_1}
    + \alpha_1^2 \left( \frac{P_{q,\tilde Z_1}}{d_\textrm{min,1}} - \frac{\sigma_{z_1}^2}{\sigma_{x_1|u_2}^2} \mathcal D_1 \right)
    \label{eq:distortion-3}
\end{align}
where $(a)$ follows by \eqref{eq:cross-correlation-1} and \eqref{eq:Zprime-variance-bound2-3}. For large $A_1,A_2$, $M_1,M_2$ we have $\sigma_{z_1}^2 \to \sigma_{x_1|u_2}^2$ and the RHS of \eqref{eq:distortion-3} becomes $\mathcal D_1$.

\subsection{Decoder}
Consider the linear minimum mean square error (MMSE) estimate of $\bm{X}$ given $\bm{Z}'$:
\begin{align}
    \E{\bm{X} (\bm{Z}')^T} Q_{\bm{Z}'}^{-1} \bm{Z}'.
    \label{eq:dsc-LMMSE-estimator}
\end{align}
We approximate by taking the limits $A_1,A_2 \to \infty$ to obtain
\begin{align}
    \hat{\bm{X}} = \begin{pmatrix} 
    \alpha_1  & \displaystyle \frac{\rho\,\sigma_{x_1}}{\sigma_{u_2}} \\ \displaystyle
    \frac{\alpha_1\, \rho\,\sigma_{x_1} \sigma_{x_2} \sigma_{\check z_2}^2}{|Q_{X_1,U_2}|} &  \displaystyle \frac{\sigma_{x_2}}{\sigma_{u_2}}
    \end{pmatrix}  \bm{Z}'.
    \label{eq:dsc-LMMSE}
\end{align}
The expression \eqref{eq:dsc-LMMSE} follows from \eqref{eq:dsc-LMMSE-estimator} by 
modeling the pair $(Z_1',Z_2')$ as uncorrelated since
\begin{align}
    \E{Z_1' Z_2'}
    & \to \E{\alpha_1 \left( X_1 - \frac{\rho\,\sigma_{x_1}}{\sigma_{u_2}} Z_2' \right) Z_2'} \nonumber \\
    & = \alpha_1 \left( \frac{\rho\,\sigma_{x_1} \sigma_{x_2}^2}{\sigma_{u_2}} - \frac{\rho\,\sigma_{x_1}}{\sigma_{u_2}} \E{(Z_2')^2} \right) \to 0.
\end{align}
Thus, $Q_{\bm{Z}'}$ is diagonal with entries $\E{(Z_1')^2}\to\sigma_{x_1|u_2}^2$ and $\E{(Z_2')^2}\to\sigma_{x_2}^2$ on the diagonal. Moreover, the second matrix column in \eqref{eq:dsc-LMMSE} has entries
\begin{align}
    \frac{\E{X_1 Z_2'}}{\E{(Z_2')^2}}
    & \to \frac{1}{\sigma_{x_2}^2}
    \E{X_1\, \left( \frac{\sigma_{x_2}}{\sigma_{u_2}} X_2 + \tilde Z_2 \right)} = \frac{\rho\,\sigma_{x_1}}{\sigma_{u_2}}
    \\
    \frac{\E{X_2 Z_2'}}{\E{(Z_2')^2}}
    & \to \frac{1}{\sigma_{x_2}^2}
    \E{X_2\, \left( \frac{\sigma_{x_2}}{\sigma_{u_2}} X_2 + \tilde Z_2 \right)} = \frac{\sigma_{x_2}}{\sigma_{u_2}}
\end{align}
while the first column has entries
\begin{align}
    \frac{\E{X_1 Z_1'}}{\E{(Z_1')^2}}
    & \to \frac{1}{\sigma_{x_1|u_2}^2}
    \E{X_1\, \alpha_1 \left( X_1 - \frac{\rho\,\sigma_{x_1}}{\sigma_{u_2}} Z_2' \right)} 
    = \alpha_1
    \\
    \frac{\E{X_2 Z_1'}}{\E{(Z_1')^2}}
    & \to \frac{1}{\sigma_{x_1|u_2}^2}
    \E{X_2\, \alpha_1 \left( X_1 - \frac{\rho\,\sigma_{x_1}}{\sigma_{u_2}} Z_2' \right)}
    \nonumber \\
    & 
    = \frac{\alpha_1\, \rho\,\sigma_{x_1} \sigma_{x_2} \sigma_{\check z_2}^2}{|Q_{X_1,U_2}|} .
\end{align}
Observe that $\hat X_1$ is the same as \eqref{eq:dsc-sc-hatX1} and the corresponding mean square error is bounded in \eqref{eq:distortion-3}.
We further compute
\begin{align}
    & \E{\big( X_2 - \hat X_2 \big)^2}
    \overset{(a)}{\to} \sigma_{x_2}^2 \left(1 -  
    \frac{\rho^2 \sigma_{x_1}^2 \sigma_{\check z_2}^4}{\sigma_{u_2}^2 |Q_{\bm{U}}|} - \frac{\sigma_{x_2}^2}{\sigma_{u_2}^2} 
    \right) 
    = \mathcal D_2
\end{align}
where step $(a)$ uses \eqref{eq:alpha1}.
We thus have the desired result.

\subsection{Short Polar Code Design and Simulations}
\label{subsec:DS-PCQ}

We design PCQ and PCQ-mod for the Gaussian source with the covariance matrix \eqref{eq:QX-example}.
Consider $(R_1,R_2)=(1,2)$ with 8-ASK for encoder 1 and 16-ASK for encoder 2, both with natural labelings. The polar codes compress strings of length $n=256$ via multilevel successive cancellation list (SCL) encoding and decoding, and list passing across levels \cite{Prinz-Yuan-ISTC18,Karakchieva-SCC19}.

\begin{figure}[!t]
\centering
\input{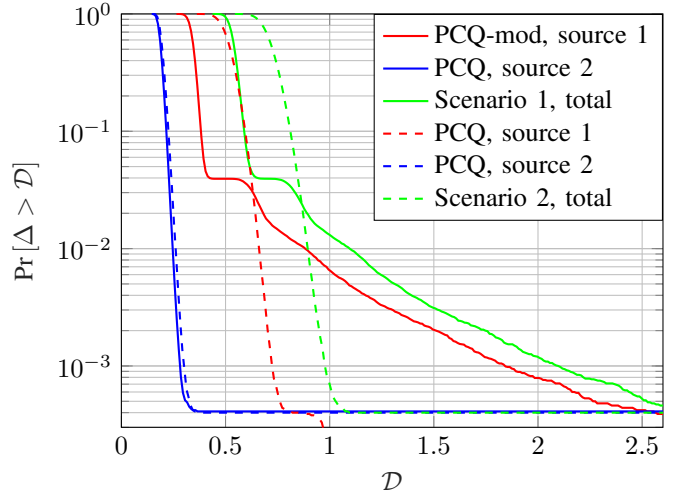}
\caption{$\Prob{\Delta>\mathcal D}$ for the source $\sigma_{x_1}^2=\sigma_{x_2}^2=2.5$ and $\rho = 0.8$, the rates $R_1=1$ and $R_2=2$, and with 8-ASK and 16-ASK for the first and second source, respectively.}
\label{fig:ds-coding}
\end{figure}

We target the corner point of the DS region, where encoder 2 uses Shannon-RD coding and encoder 1 uses WZ coding. Consider two sets of parameters for the sources $l=1,2$:
\begin{itemize} \itemsep 0pt
    \item the modulation parameters $M_l$, $\kappa_l$;
    \item the sampled Gaussian shaping parameters $\sigma_{d,l}^2$.
\end{itemize}
We optimize these parameters by targeting the ideal rates of Sec.~\ref{subsec:dsc-successive} but back off from the ideal distortions.
Fig.~\ref{fig:ds-coding} plots $\Prob{\Delta_l>\mathcal D}$ for two cases.
\begin{itemize}[leftmargin=*]
\item Case 1: WZ-PCQ-mod for source 1 and PCQ for source 2
    \begin{itemize} \itemsep 0pt
    \item WZ-PCQ-mod: $\E{\Delta_1}=0.372$ for $\kappa_1 = 1.325$ 
    \item PCQ: $\E{\Delta_2}=0.198$ for $\kappa_2 = 0.442$
    \item $\E{\Delta_1}+\E{\Delta_2} = 0.570$
    \end{itemize}
\item Case 2: PCQ and linear MMSE estimation at the decoder
\begin{itemize} \itemsep 0pt
    \item PCQ: $\E{\Delta_1}=0.531$ for $\kappa_1=0.6$
    \item PCQ: $\E{\Delta_2}=0.205$ for $\kappa_2=0.442$
    \item $\E{\Delta_1}+\E{\Delta_2} = 0.736$.
    \end{itemize}
\end{itemize}
The markers in Fig.~\ref{fig:BT-region} show the distortion pairs. WZ coding of $X_1$ significantly reduces $\E{\Delta_1}$ because the relatively large $R_2=2$ lets the decoder output an accurate representation $U_2$ that serves as strong side information to compress $X_1$. Also, $\E{\Delta_2}$ is slightly better for Case 1 than Case 2 because the reconstruction is different: Case 1 has $\hat{\bm{X}}=\E{\bm{X}| Z_1',U_2}$ and Case 2 has $\hat{\bm{X}}=\E{\bm{X}|U_1,U_2}$. 

\section{Conclusions}
\label{sec:conclusions}
We demonstrated that DS coding, based on scalar quantization, a modulo interval, dithering, and probabilistic shaping, achieves all points in the Berger-Tung region for the quadratic Gaussian problem. Simulations with short-block-length multilevel 5G polar codes illustrate the theory. Future work will apply our framework to cloud radio access networks (C-RANs) to investigate the impact of WZ coding on end-to-end information rates. Additionally, a short proof that binary multilevel coding with PCQ and PCQ-mod achieves the desired information rates seems possible by adapting the PCM proof steps in \cite{Runge-ISIT22} to the quantization setting.




\section*{Acknowledgment}
This work was supported by the German Research Foundation (DFG) via the German-Israeli Project Cooperation (DIP) under projects KR 3517/13-1 and SH 1937/1-1.

\bibliographystyle{IEEEtran}
\bibliography{references}

@article{arikan2009channel,
  title={Channel polarization: A method for constructing capacity-achieving codes for symmetric binary-input memoryless channels},
  author={Arikan, Erdal},
  journal={IEEE Trans. Inf. Theory},
  volume={55},
  number={7},
  pages={3051--3073},
  year={2009},
  publisher={IEEE}
}

@incollection{Berger-78,
  title = "Multiterminal source coding",
  booktitle = "The Information Theory Approach to Communications, ser. {CISM} Courses and Lectures",
  author = {Berger, Toby},
  pages={171–231},
  year={1978},
  editor = "Longo, G.",
  publisher={Springer Verlag},
  address={Vienna/New York}
}

@ARTICLE{Berger-Housewright-IT79,
  author={Berger, T. and Housewright, K. and Omura, J. and Suiyin Yung and Wolfowitz, J.},
  journal={IEEE Trans. Inf. Theory}, 
  title={An upper bound on the rate distortion function for source coding with partial side information at the decoder}, 
  year={1979},
  volume={25},
  number={6},
  pages={664-666},
  doi={10.1109/TIT.1979.1056105}}

@INPROCEEDINGS{Boecherer-WC17,
  author={Böcherer, Georg and Prinz, Tobias and Yuan, Peihong and Steiner, Fabian},
  booktitle={IEEE Wireless Commun. Networking Conf. Workshops}, 
  title={Efficient Polar Code Construction for Higher-Order Modulation}, 
  year={2017},
  address={San Francisco, CA, USA},
  volume={},
  number={},
  pages={1-6},
  doi={10.1109/WCNCW.2017.7919039}}

@ARTICLE{Boehnke-COMML20,
  author={Böhnke, Ronald and \.{I}{\c{s}}can, Onurcan and Xu, Wen},
  journal={IEEE Commun. Lett.}, 
  title={Multi-Level Distribution Matching}, 
  year={2020},
  volume={24},
  number={9},
  pages={2015-2019},
  doi={10.1109/LCOMM.2020.2993929}}

@ARTICLE{Campello-IT19,
  author={Campello, Antonio and Dadush, Daniel and Ling, Cong},
  journal={IEEE Trans. Inf. Theory}, 
  title={{AWGN}-Goodness Is Enough: Capacity-Achieving Lattice Codes Based on Dithered Probabilistic Shaping}, 
  year={2019},
  volume={65},
  number={3},
  pages={1961-1971},
  doi={10.1109/TIT.2018.2875004}}

@ARTICLE{Chen-Berger-IT08,
  author={Chen, Jun and Berger, Toby},
  journal={IEEE Trans. Inf. Theory}, 
  title={Successive {W}yner–{Z}iv Coding Scheme and Its Application to the Quadratic {G}aussian {CEO} Problem}, 
  year={2008},
  volume={54},
  number={4},
  pages={1586-1603},
  doi={10.1109/TIT.2008.917687}}

@article{Costa-IT83,
  title={Writing on dirty paper},
  author={Costa, Max},
  journal={IEEE Trans. Inf. Theory},
  volume={29},
  number={3},
  pages={439--441},
  year={1983}
}

@ARTICLE{Dongbo-IEEEA21,
  author={Dongbo, Huang and Gang, Hua and Yonggang, Xu and Hongsheng, Yin},
  journal={IEEE Access}, 
  title={Nested Lattice Coding With Algebraic Encoding and Geometric Decoding}, 
  year={2021},
  volume={9},
  number={},
  pages={11598-11609},
}

@INPROCEEDINGS{Eghbalian-Arani-ISWCS13,
  author={Eghbalian-Arani, Sajjad and Behroozi, Hamid},
  booktitle={Int. Symp. Wireless Commun. Sys.}, 
  title={Polar Codes for a Quadratic-{G}aussian {W}yner-{Z}iv Problem},
  address={Ilmenau, Germany},
  year={2013},
  volume={},
  number={},
  pages={1-5}
}

@Book{ElGamal-Kim-11,
  author     = "A. El Gamal and Y.-H. Kim",
  title          = "Network Information Theory",
  publisher="Cambridge Univ. Press",
  year         = 2011,
}

@ARTICLE{Honda-Yamamoto-IT13,
  author={Honda, Junya and Yamamoto, Hirosuke},
  journal={IEEE Trans. Inf. Theory}, 
  title={Polar coding Without alphabet extension for asymmetric models}, 
  year={2013},
  volume={59},
  number={12},
  pages={7829-7838},
  doi={10.1109/TIT.2013.2282305}}

@ARTICLE{Imai:IT77,
  author={Imai, H. and Hirakawa, S.},
  journal={IEEE Trans. Inf. Theory}, 
  title={A new multilevel coding method using error-correcting codes}, 
  year={1977},
  volume={23},
  number={3},
  pages={371-377},
  doi={10.1109/TIT.1977.1055718}}

@ARTICLE{Iscan-COMM18,
  author={\.{I}{\c{s}}can, Onurcan and Böhnke, Ronald and Xu, Wen},
  journal={IEEE Commun. Lett.}, 
  title={Shaped Polar Codes for Higher Order Modulation}, 
  year={2018},
  volume={22},
  number={2},
  pages={252-255},
  doi={10.1109/LCOMM.2017.2766621}}

@ARTICLE{Iscan-IA19,
  author={\.{I}{\c{s}}can, Onurcan and Böhnke, Ronald and Xu, Wen},
  journal={IEEE Access}, 
  title={Probabilistic Shaping Using {5G New Radio} Polar Codes}, 
  year={2019},
  volume={7},
  number={},
  pages={22579-22587},
  doi={10.1109/ACCESS.2019.2898103}}

@article{Iscan-TETT20,
author = {\.{I}{\c{s}}can, Onurcan and Böhnke, Ronald and Xu, Wen},
title = {Sign-bit shaping using polar codes},
journal = {Trans. Emerging Telecommun. Technol.},
volume = {31},
number = {10},
pages = {e4058},
doi = {https://doi.org/10.1002/ett.4058},
year = {2020}
}

@ARTICLE{Jha-JSAIT22,
  author={Jha, Shubham},
  journal={IEEE J. Sel. Areas Inf. Theory}, 
  title={Universal {G}aussian Quantization With Side-Information Using Polar Lattices}, 
  year={2022},
  volume={3},
  number={4},
  pages={639-650},
  doi={10.1109/JSAIT.2023.3247864}}

@INPROCEEDINGS{Karakchieva-SCC19,
  author={Karakchieva, Liudmila and Trifonov, Peter},
  booktitle={Int. ITG Conf. Sys., Commun. Coding}, 
  title={Joint list multistage decoding with sphere detection for polar coded {SCMA} systems}, 
  year={2019},
  address={Rostock, Germany},
  volume={},
  number={},
  pages={1-6},
}

@incollection{Kim-Kramer-22,
  title = "Information theory for cellular wireless networks",
  booktitle = "Information Theoretic Perspectives on 5{G} Systems and Beyond",
  author = {Kim, Y.-H. and Kramer, G.},
  pages={10–92},
  year={2022},
  editor = {Mari\'c, I. and Shamai, S. and Simeone, O.},
  publisher={Cambridge Univ. Press}
}

@ARTICLE{Korada-IT10,
  author={Korada, Satish Babu and Urbanke, Rüdiger L.},
  journal={IEEE Trans. Inf. Theory}, 
  title={Polar codes are optimal for lossy source coding}, 
  year={2010},
  volume={56},
  number={4},
  pages={1751-1768},
  doi={10.1109/TIT.2010.2040961}}

@ARTICLE{Liu-COMM16,
  author={Liu, Ling and Ling, Cong},
  journal={IEEE Trans. Commun.}, 
  title={Polar Codes and Polar Lattices for Independent Fading Channels}, 
  year={2016},
  volume={64},
  number={12},
  pages={4923-4935},
  doi={10.1109/TCOMM.2016.2613109}}

@ARTICLE{Liu-COMM19,
  author={Liu, Ling and Yan, Yanfei and Ling, Cong and Wu, Xiaofu},
  journal={IEEE Trans. Commun.}, 
  title={Construction of Capacity-Achieving Lattice Codes: Polar Lattices}, 
  year={2019},
  volume={67},
  number={2},
  pages={915-928},
  doi={10.1109/TCOMM.2018.2876113}}

@ARTICLE{Liu-IT21,
  author={Liu, Ling and Shi, Jinwen and Ling, Cong},
  journal={IEEE Trans. Inf. Theory}, 
  title={Polar Lattices for Lossy Compression}, 
  year={2021},
  volume={67},
  number={9},
  pages={6140-6163},
  doi={10.1109/TIT.2021.3097965}}

@INPROCEEDINGS{Liu-ISIT24,
  author={Liu, Ling and Lyu, Shanxiang and Ling, Cong and Bai, Baoming},
  booktitle={IEEE Int. Symp. Inf. Theory}, 
  title={On the Equivalence Between Probabilistic Shaping and Geometric Shaping: A Polar Lattice Perspective}, 
  year={2024},
  address={Athens, Greece},
  volume={},
  number={},
  pages={2174-2179},
  doi={10.1109/ISIT57864.2024.10619275}}

@ARTICLE{Mondelli-IT18,
  author={Mondelli, Marco and Hassani, S. Hamed and Urbanke, Rüdiger L.},
  journal={IEEE Trans. Inf. Theory}, 
  title={How to Achieve the Capacity of Asymmetric Channels}, 
  year={2018},
  volume={64},
  number={5},
  pages={3371-3393},
  doi={10.1109/TIT.2018.2789885}}

@INPROCEEDINGS{Prinz-SPAWC17,
  author={Prinz, Tobias and Yuan, Peihong and Böcherer, Georg and Steiner, Fabian and \.{I}{\c{s}}can, Onurcan and Böhnke, Ronald and Xu, Wen},
  booktitle={IEEE Int. Workshop Signal Proc. Advances in Wireless Commun.}, 
  title={Polar coded probabilistic amplitude shaping for short packets}, 
  year={2017},
  address={Sapporo, Japan},
  volume={},
  number={},
  pages={1-5},
  doi={10.1109/SPAWC.2017.8227653}}

@INPROCEEDINGS{Prinz-Yuan-ISTC18,
  author={Prinz, Tobias and Yuan, Peihong},
  booktitle={IEEE Int. Symp. Turbo Codes Iter. Inf. Proc.}, 
  title={Successive Cancellation List Decoding of {BMERA} Codes with Application to Higher-Order Modulation}, 
  year={2018},
  address={Hong Kong, China},
  volume={},
  number={},
  pages={1-5},
  doi={10.1109/ISTC.2018.8625293}}

@INPROCEEDINGS{Runge-ISIT22,
  author={Runge, Constantin and Wiegart, Thomas and Lentner, Diego and Prinz, Tobias},
  booktitle={IEEE Int. Symp. Inf. Theory}, 
  title={Multilevel Binary Polar-Coded Modulation Achieving the Capacity of Asymmetric Channels}, 
  year={2022},
  address={Espoo,Finland},
  volume={},
  number={},
  pages={2595-2600},
  doi={10.1109/ISIT50566.2022.9834740}}

@INPROCEEDINGS{Runge-ISIT24,
  author={Runge, Constantin and Kramer, Gerhard},
  booktitle={IEEE Int. Symp. Inf. Theory}, 
  title={Time-Shifted Alternating {G}elfand-{P}insker Coding for Broadcast Channels}, 
  year={2024},
  address={Athens, Greece},
  volume={},
  number={},
  pages={1700-1705},
  doi={10.1109/ISIT57864.2024.10619498}}

@ARTICLE{Seidl-IT13,
  author={Seidl, Mathis and Schenk, Andreas and Stierstorfer, Clemens and Huber, Johannes B.},
  journal={IEEE Trans. Commun.}, 
  title={Polar-coded modulation}, 
  year={2013},
  volume={61},
  number={10},
  pages={4108-4119},
  doi={10.1109/TCOMM.2013.090513.130433}}

@article{Sener-CL24,
  title={Achieving the Dirty Paper Channel Capacity With Scalar
  Lattices and Probabilistic Shaping},
  author={{\c{S}}ener, M Yusuf and Böhnke, Ronald and Xu, Wen and Kramer, Gerhard},
  journal={IEEE Commun. Lett.},
  volume={28},
  number={1},
  pages={29--33},
  year={2024}
}

@INPROCEEDINGS{Sener-ISIT24,
  author={{\c{S}}ener, M. Yusuf and Kramer, Gerhard and Shamai Shitz, Shlomo and Böhnke, Ronald and Xu, Wen},
  booktitle={IEE Int. Symp. Inf. Theory}, 
  title={Achieving {G}aussian Vector Broadcast Channel Capacity with Scalar Lattices}, 
  year={2024},
  address={Athens, Greece},
  volume={},
  number={},
  pages={1706-1711},
  doi={10.1109/ISIT57864.2024.10619389}}

@INPROCEEDINGS{Sener-ISIT25,
  author={Şener, M. Yusuf and Kramer, Gerhard and Shitz, Shlomo Shamai and Xu, Wen},
  booktitle={IEE Int. Symp. Inf. Theory},
  title={Scalar Lattices and Probabilistic Shaping for Dithered Wyner-Ziv Quantization}, 
  year={2025},
  address={Ann Arbor, MI, USA},
  volume={},
  number={},
  pages={1-6},
  doi={10.1109/ISIT63088.2025.11195245}}

@article{Sener-CL21,
  title={Dirty paper coding based on polar codes and probabilistic shaping},
  author={{\c{S}}ener, M Yusuf and Böhnke, Ronald and Xu, Wen and Kramer, Gerhard},
  journal={IEEE Commun. Lett.},
  volume={25},
  number={12},
  pages={3810--3813},
  year={2021},
  publisher={IEEE}
}

@INPROCEEDINGS{Sutter-ITW12,
  author={Sutter, David and Renes, Joseph M. and Dupuis, Frédéric and Renner, Renato},
  booktitle={IEEE Inf. Theory Workshop}, 
  title={Achieving the capacity of any {DMC} using only polar codes}, 
  year={2012},
  address={Lausanne, Switzerland},
  volume={},
  number={},
  pages={114-118},
  doi={10.1109/ITW.2012.6404638}}

@phdthesis{Tung-thesis-78,
  author = {Tung, S. Y.},
  title = "Multiterminal source coding",
  school = "School of Elect. Eng., Cornell Univ.",
  address = "Ithaca, NY, USA",
  year    = 1978,
  doi = {10.5075/epfl-thesis-4461}
}

@article{wachsmann1999multilevel,
  title={Multilevel codes: theoretical concepts and practical design rules},
  author={Wachsmann, Udo and Fischer, Robert F.H. and Huber, Johannes B.},
  journal={IEEE Trans. Inf. Theory},
  volume={45},
  number={5},
  pages={1361--1391},
  year={1999},
}

@ARTICLE{Wiegart-CL19,
  author={Wiegart, Thomas and Steiner, Fabian and Schulte, Patrick and Yuan, Peihong},
  journal={IEEE Commun. Lett.}, 
  title={Shaped On-Off Keying Using Polar Codes}, 
  year={2019},
  volume={23},
  number={11},
  pages={1922-1926},
  doi={10.1109/LCOMM.2019.2930511}}

@ARTICLE{Wyner-Ziv-IT76,
  author={Wyner, A. and Ziv, J.},
  journal={IEEE Trans. Inf. Theory}, 
  title={The rate-distortion function for source coding with side information at the decoder}, 
  year={1976},
  volume={22},
  number={1},
  pages={1-10},
  doi={10.1109/TIT.1976.1055508}}

@book{zamir14,
  title={Lattice Coding for Signals and Networks},
  author={Zamir, R.},
  year={2014},
  publisher={Cambridge Univ. Press},
  address={Cambridge, U.K.}
}

@article{zamir2002nested,
  title={Nested linear/lattice codes for structured multiterminal binning},
  author={Zamir, Ram and Shamai, Shlomo and Erez, Uri},
  journal={IEEE Trans. Inf. Theory},
  volume={48},
  number={6},
  pages={1250--1276},
  year={2002},
  publisher={IEEE}
}

@INPROCEEDINGS{Zamir-Shamai-ISIT98,
  author={Zamir, R. and Shamai, S.},
  booktitle={Inf. Theory Workshop}, 
  title={Nested linear/lattice codes for {W}yner-{Z}iv encoding}, 
  year={1998},
  address={Killarney, Ireland},
  volume={},
  number={},
  pages={92-93},
  doi={10.1109/ITW.1998.706450}}

@article{wagner2008rate,
  title={Rate region of the quadratic {G}aussian two-encoder source-coding problem},
  author={Wagner, Aaron B and Tavildar, Saurabha and Viswanath, Pramod},
  journal={IEEE Trans. Inf. Theory},
  volume={54},
  number={5},
  pages={1938--1961},
  year={2008},
  publisher={IEEE}
}

@article{oohama2005rate,
  title={Rate-distortion theory for {G}aussian multiterminal source coding systems with several side informations at the decoder},
  author={Oohama, Yasutada},
  journal={IEEE Trans. Inf. Theory},
  volume={51},
  number={7},
  pages={2577--2593},
  year={2005},
  publisher={IEEE}
}

@inproceedings{prabhakaran2004rate,
  title={Rate region of the quadratic {G}aussian {CEO} problem},
  author={Prabhakaran, Vinod and Tse, David and Ramachandran, Kannan},
  booktitle={IEEE Int. Symp. Inf. Theory},
address={Chicago, IL},
  pages={119},
  year={2004},
  organization={IEEE}
}

@article{berger1996ceo,
  title={The {CEO} problem [multiterminal source coding]},
  author={Berger, Toby and Zhang, Zhen and Viswanathan, Harish},
  journal={IEEE Trans. Inf. Theory},
  volume={42},
  number={3},
  pages={887--902},
  year={1996},
  publisher={IEEE}
}

@phdthesis{ghaddar2022channel,
  author={Ghaddar, Nadim},
  title={Channel Coding Techniques for Communication over Networks and over Channels with Memory},
  school={University of California, San Diego},
  year={2022},

}

\clearpage
\appendices
\end{document}